\documentstyle{article}
\title{
\begin{flushright}
{\bf\normalsize   COLO-HEP-316 }  \\
\end{flushright}
\vskip 10pt
\bf Freezing a Fluid Random Surface
}

\author{ {\it C.F. Baillie} \\
         Physics Dept. \\
         University of Colorado\\
         Boulder, CO 80309\\
         USA\\
         \\
         {\it D.A. Johnston} \\
         Dept. of Mathematics\\
         Heriot-Watt University\\
         Riccarton\\
         Edinburgh, EH14 4AS\\
         Scotland}
\begin{document}
  \maketitle
                      {\Large
                      \begin{abstract}
%
We investigate a dynamically triangulated random surface action
that consists of a gaussian term plus the modulus of the intrinsic scalar
curvature. We find that the flips are frozen out and the internal geometry
is regularized as the coefficient of the latter term is increased. Such a term
thus provides a way of interpolating between dynamically triangulated
(ie fluid) and crystalline random surfaces.
%
                        \end{abstract} }
%
  \thispagestyle{empty}
%
%
  \newpage
%
                  \pagenumbering{arabic}
The partition function for a dynamically triangulated random surface
incorporates
a sum over triangulations (or more generally, polygonizations) of the surface
$\sum_{T}$
which means that we are dealing with a fluid surface:
\begin{equation}
Z = \sum_{T} \int \prod_{i=1}^{N-1} d X_i^{\mu} \exp (- S_g).
\label{e2}
\end{equation}
If we were motivated by investigating string theory $S_g$ could be a simple
gaussian term,
\begin{equation}
S_g = {1 \over 2} \sum_{<ij>} (X_i^{\mu} - X_j^{\mu})^2,
\label{e3}
\end{equation}
arising from the discretized Polyakov string action \cite{1},
where the $X$'s live at the vertices of the triangulation and the sum $<ij>$ is
over all the edges.
It is also possible to consider area and edge length actions \cite{2,3} in this
context, but all
fail to lead to a sensible continuum theory because the string
tension does not scale \cite{4}.

There have also been numerous simulations and analytical treatments of
fixed (triangulation) or crystalline random surfaces in which the sum over
triangulations is dropped in favour of a single random or regular
discretization
of the surface. These can be thought of either as models in their own right of
polymerized membranes
in solid state physics or as an approximation to the dynamically triangulated
surfaces
which hopefully retain some of the salient behaviour. Various actions, many
incorporating a
``stiffness'' or extrinsic curvature term \cite{5,5a}
\begin{equation}
S_e = \sum_{\Delta_i, \Delta_j} ( 1 - n_i \cdot n_j),
\label{e4}
\end{equation}
where $n_i, n_j$ are the normals on neighboring triangles $\Delta_i, \Delta_j$,
have been looked at in the solid state physics context. Initial simulations
\cite{6a} suggested
that there was a second order
phase transition for fixed surfaces (the ``crumpling transition'') as the
coefficient of $S_e$ was increased,
though later work has shown that the incorporation of realistic features such
as self-avoidance suppresses this \cite{6b,6c}.
It is clear, however, that the crumpling transition does continue to exist in
the absence of self-avoidance on a fixed
surface and is second order \cite{9}.
The existence of this transition on a fixed surface has spurred many groups to
investigate actions incorporating
stiffness terms on dynamically triangulated surfaces in the hope of finding a
viable continuum limit
for a string theory \cite{6}. The initial simulations suggested a second order
transition, just as for a fixed surface,
but more recent simulations of larger surfaces
indicate a higher order transition \cite{7,8} at which the string tension may,
however, still be scaling
satisfactorily. Analytical work is not supportive of the existence of a
transition for dynamically triangulated
surfaces \cite{5}, suggesting that
the coupling of the stiffness is asymptotically free, and this is backed up by
the calculations for fixed
surfaces where the in-plane rigidity mediates the transition. The calculations
for the dynamically
triangulated surfaces do miss out the Liouville mode and it is conceivable that
this plays an important role
in a transition if one exists, by inducing long range correlations.
The status of the transition with alternative actions, such as the ``Steiner''
term suggested by Savvidy et.al. is
also rather uncertain \cite{10}.

In view of the uncertainties surrounding the behaviour of dynamically
triangulated surfaces it would
be useful to have some means of interpolating between fixed or crystalline
surfaces, where the crumpling
transition is better understood, and dynamically triangulated surfaces. A
regular, flat triangulation
will have each vertex of the triangulation surrounded by six others so a term
in the action of the form
\begin{equation}
S \simeq \sum_i (q_i - 6 )
\label{e5}
\end{equation}
where $q_i$ is the number of neighbours of vertex $i$,
immediately springs to mind as a means of smoothing the
internal geometry, but this is proportional to the Euler characteristic of the
surface and
will not contribute to the dynamics.  Similarly, we can try
\begin{equation}
S \simeq \sum_i  (q_i - 6)^2
\label{e6}
\end{equation}
but this is the discretization
of a higher derivative $R^2$ term and would not
be expected to contribute in the continuum limit because it is irrelevant on
dimensional
grounds \footnote{On very small meshes
a term of the form $\sum_i \log (q_i)$ which generates an $R^2$ if expanded
{\it does}
appear to affect the surface configurations if it is included with a very large
coefficient, but this is almost certainly a finite size effect \cite{11}.}.
An action of this form has been used to simulate the statistical mechanics and
dynamics
of unembedded Voronoi lattices \cite{11a} where it was found that
it did have a smoothing effect on the geometry, but that there was no sign of a
peak in the associated
specific heat.
A third possibility is to consider
\begin{equation}
S_{freeze} = {\pi \over 3} \sum_i | q_i - 6 |
\label{e7}
\end{equation}
where the modulus sign ensures that we do not get the Euler characteristic
and the normalization is chosen to agree with  \cite{8}
where the term was used as an observable rather than being included
in the dynamics. This term is non-topological and
will not be suppressed by inverse powers of the cutoff, as equ.(\ref{e6}) is,
because its dimension
is lower.

$S_{freeze}$ is also a natural object to consider from a geometrical point of
view. For a curve $C$
embedded in three dimensions it was shown by Fenchel that
\begin{equation}
{1 \over \pi} \int_C | \kappa | ds \ge 2
\label{e8}
\end{equation}
where $\kappa$ is the curvature and the equality (the case of a ``tight
immersion'')
holds when $C$ is a plane convex curve \cite{15}.
In \cite{12} we investigated a version of this for surfaces
\begin{equation}
S_{tight} = \sum_i | 2 \pi   - \sum_{j(i)}  \phi_{ij}  |
\label{e9}
\end{equation}
where the outer sum is over all the vertices of the triangulation and the inner
sum is round the
neighbors
$j$ of a node $i$. $\phi_{ij}$ is the angle subtended by the $j$th triangle at
the $i$th vertex.
$S_{tight}$ would be maximized for convex immersions
and might therefore be expected to help smooth
out surfaces, but we found that it was insufficient to overcome the disordering
effect of area or gaussian
terms. $S_{freeze}$ is the equivalent of $S_{tight}$ for the {\it intrinsic}
geometry of the surface and
in what follows we will examine the effect of this term on both the intrinsic
and extrinsic geometry
of a dynamically triangulated random surface.
For compatibility with our earlier work we include the contribution of the
measure in the action
\begin{equation}
S_m =  - { d \over 2 } \sum_i \log ( q_i ),
\label{e10}
\end{equation}
and simulate $S_g + S_m + \lambda S_{freeze}$ for various $\lambda$.

As we are varying the coefficient of a term in the intrinsic geometry,
$S_{freeze}$,
we would expect most of the effects to be intrinsic rather than extrinsic.
We therefore measure $S_{freeze}$ itself and the variance (or specific heat)
term associated with it
\begin{equation}
C = {\lambda^2 \over N} \left( < S_{freeze}^2 > - < S_{freeze} >^2 \right).
\label{e11}
\end{equation}
We also measure the mean maximum number of neighbours $\max(q_i)$
and the measure term $S_m$ to get an impression of the
irregularity of the internal geometry as $\lambda$ is varied.
A useful check that the simulation is performing as it should is
provided by removing the modulus sign in equ.(\ref{e7}) which should then sum
to give $4 \pi$ for
every configuration by the Gauss-Bonnet theorem.
The extrinsic geometry is monitored by looking at the gyration radius $X2$
\begin{equation}
X2 = { 1 \over 9 N (N -1)} \sum_{ij} \left( X^\mu_i - X^\mu_j \right)^2
q_i q_j.
\label{e12}
\end{equation}
which is small in a crumpled phase and large in a smooth phase,
as well as by direct inspection of ``snapshots'' of the surfaces.
The expectation value of $S_g$ can be shown to be $d( N-1)/ 2$
for all $\lambda$
by exactly the same scaling argument that is used to give this result for the
gaussian plus
stiffness actions, in this case because
$S_{freeze}$ is independent of $X$ and therefore trivially scale invariant.
This serves as an additional useful check of equilibration.

The simulation used a Monte Carlo procedure
which we have described in some detail elsewhere \cite{16}. It first goes
through the mesh moving the $X$'s, carrying out a Metropolis
accept/reject at each step, and then goes through the mesh again
carrying out the ``flip'' moves on the links, again applying a
Metropolis accept/reject at each stage. The entire procedure
constitutes a sweep. Due to the correlated nature of the
data, a measurement was taken every tenth sweep and binning
techniques were used to analyze the errors. We carried out 10K
thermalization sweeps  followed by 30K  measurement sweeps for
each data point for the 72 node surfaces and 50K measurement sweeps for the
144 node surfaces. The acceptance for the $X$ move
was monitored and the size of the shift was adjusted to maintain an
acceptance of around 50 percent. The acceptance for the flip move
was also measured, but this can not be adjusted. As we shall see below
it is this measurement that provides the greatest interest.

\vskip 10pt
\hoffset=0truein
\voffset=0truein
\centerline{
\hbox{\vbox{ \tabskip=0pt \offinterlineskip
\def\tablerule{\noalign{\hrule}}
\halign{\strut#& \vrule#&
\hfill#\hfill &\vrule#& \hfill#\hfill &\vrule#&
\hfill#\hfill &\vrule#& \hfill#\hfill &\vrule#&
\hfill#\hfill &\vrule#& \hfill#\hfill &\vrule#&
\hfill#\hfill &\vrule#& \hfill#\hfill &\vrule#&
\hfill#\hfill &\vrule#
\tabskip=0pt\cr
\tablerule
\omit&height2pt& \omit&height2pt&
\omit&height2pt& \omit&height2pt&
\omit&height2pt& \omit&height2pt&
\omit&height2pt& \omit&height2pt&
\omit&height2pt&
\omit&\cr
&&\enskip $\lambda$ \enskip
&&\enskip sweeps \enskip
&&\enskip $S_g$ \enskip
&&\enskip $S_m$ \enskip
&&\enskip $S_{freeze}$ \enskip
&&\enskip $C$ \enskip
&&\enskip $X2$ \enskip
&&\enskip $max(q_i)$ \enskip
&&\enskip $Fs$ \enskip
&\cr
\omit&height2pt& \omit&height2pt&
\omit&height2pt& \omit&height2pt&
\omit&height2pt& \omit&height2pt&
\omit&height2pt& \omit&height2pt&
\omit&height2pt&
\omit&\cr
\tablerule
\omit&height2pt& \omit&height2pt&
\omit&height2pt& \omit&height2pt&
\omit&height2pt& \omit&height2pt&
\omit&height2pt& \omit&height2pt&
\omit&height2pt&
\omit&\cr
&& 0.500 && 30K &&   106.38(0.01) &&   123.63(0.00) &&    99.13(0.01) &&
0.37(  0.00) &&     1.94(0.01) &&    11.64(0.00)
&&0.47 &\cr
&& 1.000 && 30K &&   106.56(0.04) &&   125.42(0.00) &&    57.76(0.01) &&
0.84(  0.00) &&     2.09(0.01) &&     9.27(0.00)
&&0.26 &\cr
&& 1.500 && 30K &&   106.48(0.06) &&   126.17(0.00) &&    35.08(0.04) &&
1.04(  0.00) &&     2.00(0.01) &&     8.06(0.00)
&&0.12 &\cr
&& 1.750 && 30K &&   106.39(0.06) &&   126.37(0.00) &&    27.80(0.03) &&
1.04(  0.00) &&     1.88(0.02) &&     7.62(0.00)
&&0.08  &\cr
&& 2.000 && 30K &&   106.25(0.04) &&   126.52(0.00) &&    22.42(0.03) &&
0.98(  0.00) &&     1.71(0.02) &&     7.30(0.00)
&&0.05  &\cr
&& 2.250 && 30K &&   106.39(0.05) &&   126.60(0.00) &&    18.86(0.07) &&
0.87(  0.01) &&     1.63(0.02) &&     7.09(0.00)
&&0.04  &\cr
&& 2.500 && 30K &&   106.62(0.05) &&   126.68(0.00) &&    16.21(0.01) &&
0.62(  0.00) &&     1.54(0.00) &&     6.88(0.00)
&&0.03  &\cr
&& 3.000 && 30K &&   106.48(0.02) &&   126.73(0.00) &&    13.91(0.02) &&
0.34(  0.01) &&     1.54(0.00) &&     6.49(0.00)
&&0.01  &\cr
\omit&height2pt& \omit&height2pt&
\omit&height2pt& \omit&height2pt&
\omit&height2pt& \omit&height2pt&
\omit&height2pt& \omit&height2pt&
\omit&height2pt&
\omit&\cr
\tablerule
}}}}
\smallskip
\centerline{Table 1}
\centerline{Results for $N=72$}
\vskip 5pt

The results for the 72 node surfaces are shown in Table 1.
If we look at $S_g$ we can see that it remains close to $d/2(N-1) = 106.5$
for all the $\lambda$ values so the results appear to be fairly well
equilibrated.
We have not tabulated the Euler characteristic obtained by dropping the modulus
sign in $S_{freeze}$
but this also comes out correctly.
The results for $X2$ show that $S_{freeze}$ does not smooth out the extrinsic
geometry
as $\lambda$ is increased indeed, if anything, it has the reverse effect --
see Fig. 1.
However, the effect of the term on the intrinsic geometry is much more marked.
We can see from the values for $max(q_i)$ that the geometry is regularized
quite rapidly as $\lambda$ is increased and that by $\lambda=3$ the surfaces
have
an essentially flat internal geometry.
Even more striking is that the flip acceptance $Fs$ drops  sharply
to essentially zero as $\lambda$ is increased. We have not seen this with
the other actions we have simulated such as $S_g + \lambda S_e$ or $S_g +
\lambda S_{tight}$,
which remain fluid at any apparent phase transitions. We have thus frozen the
internal
geometry by the addition of $S_{freeze}$.

\vskip 10pt
\hoffset=0truein
\voffset=0truein
\centerline{
\hbox{\vbox{ \tabskip=0pt \offinterlineskip
\def\tablerule{\noalign{\hrule}}
\halign{\strut#& \vrule#&
\hfill#\hfill &\vrule#& \hfill#\hfill &\vrule#&
\hfill#\hfill &\vrule#& \hfill#\hfill &\vrule#&
\hfill#\hfill &\vrule#& \hfill#\hfill &\vrule#&
\hfill#\hfill &\vrule#& \hfill#\hfill &\vrule#&
\hfill#\hfill &\vrule#
\tabskip=0pt\cr
\tablerule
\omit&height2pt& \omit&height2pt&
\omit&height2pt& \omit&height2pt&
\omit&height2pt& \omit&height2pt&
\omit&height2pt& \omit&height2pt&
\omit&height2pt&
\omit&\cr
&&\enskip $\lambda$ \enskip
&&\enskip sweeps \enskip
&&\enskip $S_g$ \enskip
&&\enskip $S_m$ \enskip
&&\enskip $S_{freeze}$ \enskip
&&\enskip $C$ \enskip
&&\enskip $X2$ \enskip
&&\enskip $max(q_i)$ \enskip
&&\enskip $Fs$ \enskip
&\cr
\omit&height2pt& \omit&height2pt&
\omit&height2pt& \omit&height2pt&
\omit&height2pt& \omit&height2pt&
\omit&height2pt& \omit&height2pt&
\omit&height2pt&
\omit&\cr
\tablerule
\omit&height2pt& \omit&height2pt&
\omit&height2pt& \omit&height2pt&
\omit&height2pt& \omit&height2pt&
\omit&height2pt& \omit&height2pt&
\omit&height2pt&
\omit&\cr
&& 0.500 && 50K &&   214.49(0.01) &&   249.11(0.00) &&   202.23(0.02) &&
0.39(  0.00) &&     2.53(0.01) &&    13.01(0.00) &&0.47 &\cr
&& 1.000 && 50K &&   214.42(0.03) &&   252.92(0.00) &&   115.57(0.04) &&
0.89(  0.00) &&     2.76(0.01) &&    10.14(0.00) &&0.27 &\cr
&& 1.250 && 50K &&   214.49(0.02) &&   253.88(0.00) &&    87.89(0.01) &&
1.03(  0.00) &&     2.93(0.02) &&     9.29(0.00) &&0.19 &\cr
&& 1.750 && 50K &&   214.38(0.07) &&   254.92(0.00) &&    52.02(0.03) &&
1.10(  0.00) &&     3.24(0.08) &&     8.20(0.00) &&0.08 &\cr
&& 2.000 && 50K &&   214.80(0.06) &&   255.19(0.00) &&    40.78(0.10) &&
1.11(  0.01) &&     3.53(0.31) &&     7.83(0.01) &&0.05 &\cr
&& 2.250 && 50K &&   214.30(0.04) &&   255.36(0.00) &&    32.20(0.12) &&
1.08(  0.00) &&     3.46(0.06) &&     7.50(0.01) &&0.03 &\cr
&& 2.500 && 50K &&   214.09(0.04) &&   255.51(0.00) &&    25.96(0.03) &&
0.93(  0.01) &&     2.85(0.08) &&     7.27(0.00) &&0.02 &\cr
&& 3.000 && 50K &&   214.78(0.07) &&   255.69(0.00) &&    17.68(0.03) &&
0.62(  0.00) &&     1.78(0.01) &&     6.97(0.00) &&0.01 &\cr
\omit&height2pt& \omit&height2pt&
\omit&height2pt& \omit&height2pt&
\omit&height2pt& \omit&height2pt&
\omit&height2pt& \omit&height2pt&
\omit&\cr
\tablerule
}}}}
\smallskip
\centerline{Table 2}
\centerline{Results for $N=144$}
\vskip 5pt

The results from the simulations of the 144 node surfaces, Table 2,
are in accordance with the above.
There is still a gentle bump in the specific heat
associated with $S_{freeze}$ which gives some indication that we may be seeing
a genuine
transition rather than a crossover, though it is difficult to say for certain
with
the modest simulations we have carried out here. The value of $S_g$ ($\simeq
214.5$ in this case)
and measurements of the Euler characteristic again indicate that the simulation
is
performing as it should. Even on these larger surfaces $S_{freeze}$ is most
effective at freezing
out the flip moves on the surface, which reassures us that we are not seeing a
finite
size effect. The value of $\max(q_i)$ ($\ne 6$)
indicates that the geometry is not
completely regular in the frozen phase. The effect of the spherical
topology is to {\it reduce} the average value of $q_i$ (an icosahedron,
for example, has $q_i=5$ for all the points) so
there must be defects of $q_i=7$
frozen in too, though there cannot be many as $S_{freeze}$ is close to
its minimum bound of $4 \pi$.

{}From these results it is clear that the addition of $S_{freeze}$ to a
gaussian action
allows us to interpolate between a dynamically triangulated surface with a
fluctuating
internal geometry and a surface with a fixed, regular ($q_i \simeq 6$)
geometry.
As this is an internal effect that appears to be largely decoupled from the
extrinsic
geometry, there is no reason to expect anything different for other actions
such as those used in hunting for continuum string theories.
An interpolation between fixed and dynamical surfaces with this method would
then be of great interest in examining
the crumpling transition as the intrinsic geometry is gradually frozen (or
unfrozen!)
to see how it changes. The intermediate state where the surface is not
completely frozen might allow the modelling of fluid
surfaces with some rigid internal structure \footnote{Such as red blood cells.}
which have previously been investigated using simulations in which a rigid
frame is explicitly included \cite{99}.
It would also be useful to examine cases such as the Ising model
where analytical results are available
to determine the nature of the crossover from fixed to dynamical critical
exponents.

This work was supported in part by NATO collaborative research grant CRG910091.
CFB is supported by DOE under contract DE-FG02-91ER40672
and by NSF Grand Challenge Applications Group Grant ASC-9217394.
The computations were performed on workstations at
Heriot-Watt University.
We would like to thank
R.D. Williams for help in developing initial versions of the dynamical
mesh code.

\vfill
\eject

\bigskip

\centerline{\bf Figure Captions}
\begin{description}
\item[Fig. 1.]
A typical surface at $\lambda=3$. There is no sign of
any smoothing in the {\it extrinsic} geometry.
\end{description}

\end{document}